\documentclass[compsoc,conference,a4paper,10pt,times]{IEEEtran}
\IEEEoverridecommandlockouts

\usepackage{colortbl}
\usepackage{tikz}  
\usepackage{cite}
\usepackage{amsmath,amssymb,amsfonts}
\usepackage{graphicx}
\usepackage{textcomp}
\usepackage{bmpsize}
\usepackage{xcolor}
\usepackage{lipsum}
\usepackage[colorlinks=true,urlcolor=black]{hyperref}
\def\BibTeX{{\rm B\kern-.05em{\sc i\kern-.025em b}\kern-.08em
    T\kern-.1667em\lower.7ex\hbox{E}\kern-.125emX}}

\usepackage{colortbl}
\usepackage{tikz}          
\usepackage{amsmath}
\usepackage{graphicx}
\usepackage{booktabs}
\usepackage{latexsym}   
\usepackage{array}      
\usepackage{multirow}
\usepackage{subcaption}
\usepackage[noend]{algpseudocode}
\usepackage{threeparttable}
\usepackage{paralist}   
\usepackage{xspace}
\usepackage{color}
\usepackage{adjustbox}
\usepackage{balance}
\usepackage{wasysym}
\usepackage{rotating}   
\usepackage{listings}
\usepackage{tabu}
\usepackage{pifont}
\usepackage{framed}
\usepackage{pifont}

\usepackage{float}

\usepackage{xpatch}
\makeatletter
\chardef\TPT@@@asteriskcatcode=\catcode`*
\catcode`*=11
\xpatchcmd{\threeparttable}
  {\TPT@hookin{tabular}}
  {\TPT@hookin{tabular}\TPT@hookin{tabu}}
  {}{}
\catcode`*=\TPT@@@asteriskcatcode
\makeatother
\usepackage{tcolorbox}
\tcbuselibrary{breakable, skins}
\tcbset{%
  label begin/.style={label={#1}},
  label end/.style={after upper=\label{#1}}
}
\newtcolorbox[%
auto counter]{mybox}[2][]{%
  enhanced jigsaw,
  breakable,
  #1}

\newcommand{\shi}{\CIRCLE\xspace}
\newcommand{\kong}{\Circle\xspace}

\RequirePackage[normalem]{ulem} 
\RequirePackage{color}\definecolor{RED}{rgb}{1,0,0}\definecolor{BLUE}{rgb}{0,0,1} 

\definecolor{wheat1}{rgb}{1.000000,0.905882,0.729412}
\definecolor{LightGray}{rgb}{0.827451,0.827451,0.827451}

\newcolumntype{a}{>{\columncolor{wheat1}}l}

\definecolor{mygreen}{rgb}{0,0.6,0}
\definecolor{mygray}{rgb}{0.5,0.5,0.5}
\definecolor{mymauve}{rgb}{0.58,0,0.82}
\definecolor{darkblue}{rgb}{0.0,0.0,0.6}
\definecolor{maroon}{RGB}{102, 0, 0}
\definecolor{Maroon}{cmyk}{0,0.87,0.68,0.32}
\definecolor{darkred}{RGB}{139, 0, 0}
\definecolor{forestgreen}{RGB}{34, 139, 34}

\lstdefinelanguage{XML}
{
  basicstyle=\ttfamily\small,   
  morestring=[b],
  moredelim=[s][\color{darkblue}]{<}{\ },
  moredelim=[s][\color{darkblue}]{</}{>},
  moredelim=[l][\color{darkblue}]{/>},
  moredelim=[l][\color{darkblue}]{>},
  morecomment=[s]{<?}{?>},
  morecomment=[s]{<!--}{-->},
  stringstyle=\color{darkred},
  identifierstyle=\color{mymauve}
}

\lstdefinestyle{customJava}{
  breaklines=true,
  keepspaces=true,
  frame=single,
  language=Java,
  showstringspaces=false,
  basicstyle=\footnotesize\ttfamily,
  keywordstyle=\color{blue},
  otherkeywords={+, getIntent},
  numbers=left,
  numbersep=5pt,
  numberstyle=\scriptsize\color{black},
  rulecolor=\color{black},
  stepnumber=1,
  tabsize=2,
  commentstyle=\itshape\color{green!40!black},
  stringstyle=\color{orange},
  emph=[1]  
  {
        do,
        try,
        new,
        catch,
        while,
        SecProvider,
        SecReceiver,
        SecService,
        SecActivity,
        SecSink,
  },
  emphstyle=[1]{\color{darkred}},
  emph=[2]  
  {
        @Override,
  },
  emphstyle=[2]{\color{purple!40!black}},
  belowskip=-1em, 
}

\newif\ifANNOYMIZE
\ANNOYMIZEfalse

\newif\ifACM
\ACMfalse 

\newcommand{\myfig}{Figure\xspace}

\ifACM
\newcommand{\mysec}{Section\xspace}
\else
\newcommand{\mysec}{\S}
\fi

\newcommand{\nameDe}{\texttt{3DNDroid}\xspace}
\newsavebox{\bigimage} 

\begin{document}
\title{Shelving it rather than Ditching it: Dynamically Debloating DEX and Native Methods of Android Applications without APK Modification
    }

\author{
  \IEEEauthorblockN{Zicheng Zhang}
  \IEEEauthorblockA{\textit{Singapore Management University} \\
  Singapore, Singapore \\
  zczhang.2020@phdcs.smu.edu.sg
  }
  \and
  \IEEEauthorblockN{Jiakun Liu}
\IEEEauthorblockA{\textit{Singapore Management University} \\
  Singapore, Singapore \\
  jkliu@smu.edu.sg}
  \and
  \IEEEauthorblockN{Ferdian Thung}
  \IEEEauthorblockA{\textit{Singapore Management University} \\
  Singapore, Singapore \\
  ferdianthung@smu.edu.sg}
  \and
  \IEEEauthorblockN{Haoyu Ma}
  \IEEEauthorblockA{\textit{Zhejiang Lab} \\
  Hangzhou, China\\
  hyma@zhejianglab.com}
  \and
  \IEEEauthorblockN{Rui Li}
    \IEEEauthorblockA{\textit{Singapore Management University} \\
  Singapore, Singapore \\
  ruili@smu.edu.sg}
  \and
   \IEEEauthorblockN{Yan Naing Tun}
    \IEEEauthorblockA{\textit{Singapore Management University} \\
  Singapore, Singapore \\
  yannaingtun@smu.edu.sg}
   \and
   \IEEEauthorblockN{Wei Minn}
    \IEEEauthorblockA{\textit{Singapore Management University} \\
  Singapore, Singapore \\
  wei.minn.2023@phdcs.smu.edu.sg}
  \and
   \IEEEauthorblockN{Lwin Khin Shar}
    \IEEEauthorblockA{\textit{Singapore Management University} \\
  Singapore, Singapore \\
  lkshar@smu.edu.sg}
  \and
  \IEEEauthorblockN{Shahar Maoz}
  \IEEEauthorblockA{\textit{Tel Aviv University} \\
  Tel Aviv, Israel\\
  maoz@cs.tau.ac.il}
  \and
  \IEEEauthorblockN{Eran Toch}
  \IEEEauthorblockA{\textit{Tel Aviv University} \\
  Tel Aviv, Israel\\
  erant@tauex.tau.ac.il}
  \and
  \IEEEauthorblockN{David Lo}
  \IEEEauthorblockA{\textit{Singapore Management University} \\
  Singapore, Singapore\\
  davidlo@smu.edu.sg}
  \and
  \IEEEauthorblockN{Joshua Wong}
  \IEEEauthorblockA{\textit{Singapore Management University} \\
  Singapore, Singapore\\
  josh.onn@gmail.com}
  \and
  \IEEEauthorblockN{Debin Gao}
  \IEEEauthorblockA{\textit{Singapore Management University} \\
  Singapore, Singapore\\
  dbgao@smu.edu.sg}
}

\maketitle

\begin{abstract}
    Today's Android developers tend to include numerous features to accommodate diverse user requirements, which inevitably leads to bloated apps.
    Yet more often than not, only a fraction of these features are frequently utilized by users, thus a bloated app costs dearly in potential vulnerabilities, expanded attack surfaces, and additional resource consumption. 
    Especially in the event of severe security incidents, users have the need to block vulnerable functionalities immediately.
    Existing works have proposed various code debloating approaches for identifying and removing features of executable components. 
    However, they typically involve static modification of files (and, for Android apps, repackaging of APKs, too), which lacks user convenience let alone undermining the security model of Android due to the compromising of public key verification and code integrity checks.

    This paper introduces \nameDe, a \textbf{D}ynamic \textbf{D}ebloating approach targeting both \textbf{D}EX and \textbf{N}ative methods in An\textbf{Droid} apps. Using an unprivileged management app in tandem with a customized Android OS, \nameDe dynamically reduces unnecessary code loading during app execution based on a pre-generated debloating schema from static or dynamic analyses. It intercepts invocations of debloated bytecode methods to prevent their interpretation, compilation, and execution, while zero-filling memory spaces of debloated native methods during code loading.
Evaluation demonstrates \nameDe's ability to debloat 187 DEX methods and 30 native methods across 55 real-world apps, removing over 10K Return-Oriented Programming (ROP) gadgets. Case studies confirm its effectiveness in mitigating vulnerabilities, and performance assessments highlight its resource-saving advantages over non-debloated apps.

\end{abstract}


\section{Introduction}
\label{sec:intro}

In modern times, mobile applications (apps) have become an indispensable part of our daily lives. 
By March 2024, Android dominates the global mobile operating system market with a 70.78\% market share~\cite{AndroidMarketshare}, making millions of Android apps available to users. 
With advancements in the performance and storage capability of Android devices, apps are growing bloated as developers incorporate more features to cater to various user needs.
For example, the Grab app~\cite{Grab} is initially designed for ride-hailing but has grown into a ``super app'' offering food delivery, payment services, insurance purchases, etc.
Regardless of the commercial motivation of such effort to add more functionalities, studies have revealed that, actually, around 80\% of features in typical software products are rarely used~\cite{80-aijaz-2020}.
Yet obliviously, users would still have to install the ``super apps'' as a whole, despite the considerable amount of unused code.

The bloated apps pose a security risk, as \emph{code not needed by users could still be exploited for code reuse attacks}. 
In fact, research showed that both native library methods and compiled DEX methods of an Android app could potentially be leveraged as ROP gadgets~\cite{raja2017return,cheng2014ropecker,hanna2013juxtapp}.
Additionally, certain vulnerabilities (take CVE-2024-32876~\cite{CVE-2024-32876} as an 
illustrative example) may persist in unused code, leaving apps vulnerable to exploitation through direct inter-app code invocation~\cite{gao2020borrowing}.
Sometimes, even when the code is needed but not actively used, users may still need to debloat it immediately in the event of severe security incidents, such as the widely known vulnerability of the log4j library~\cite{CVE-2021-44832}, which allowed remote code execution and impacted all apps utilizing its logging functionality.
Moreover, even if library developers fix a vulnerability, app developers may not update their apps promptly, leaving them vulnerable. 
Previous studies~\cite{derrKeepMeUpdated2017,huangCrashEvaluatingThirdParty2019,salzaDevelopersUpdateThirdparty2018,salza2020third} revealed that only about 15\% of libraries are consistently updated by app developers.
In such cases, aside from the app developers, end users have the demand to temporarily disable vulnerable features of the app until patches are applied if they wish to continue using the app.
Lastly, the unused code also consumes additional system resources, such as memory and CPU usage~\cite{combining-bhattacharya-2013}. 



\begin{table*}[t!]
\centering
\caption{Comparison between \nameDe and existing works}
\label{tab:comparison}
\resizebox{\textwidth}{!}{
\setlength{\tabcolsep}{1mm}{
\begin{tabular}{c|ccccccc}
\hline
                                                 & \multirow{2}{*}{\begin{tabular}[c]{@{}c@{}}No modification\\ on APKs\end{tabular}} & \multirow{2}{*}{\begin{tabular}[c]{@{}c@{}}Easy to recover\\ debloated code\end{tabular}} & \multirow{2}{*}{\begin{tabular}[c]{@{}c@{}}Change debloating\\ schema at runtime\end{tabular}} & \multicolumn{3}{c}{Debloating range}                                            & \multirow{2}{*}{\begin{tabular}[c]{@{}c@{}}Method-level\\ debloating\end{tabular}} \\ \cline{5-7}
\multicolumn{1}{l|}{}                            &                                                                                    &                                                                                           &                                                                                                & \multicolumn{1}{l}{DEX} & \multicolumn{1}{l}{Native} & \multicolumn{1}{l}{Both} &                                                                                    \\ \hline
RedDroid~\cite{jiang_reddroid_2018}              & \kong                                                                              & \kong                                                                                     & \kong                                                                                          & \kong                   & \shi                       & \kong                    & \kong                                                                              \\
AutoDebloater~\cite{liu2023autodebloater}        & \kong                                                                              & \kong                                                                                     & \kong                                                                                          & \shi                    & \kong                      & \kong                    & \shi                                                                               \\
MiniMon~\cite{liu2024minimon}                    & \kong                                                                              & \kong                                                                                     & \kong                                                                                          & \shi                    & \kong                      & \kong                    & \shi                                                                               \\
Dynamic Binary Shrinking~\cite{pilgun_dont_2020} & \kong                                                                              & \kong                                                                                     & \kong                                                                                          & \shi                    & \kong                      & \kong                    & \shi                                                                               \\
XDebloat~\cite{xdebloat-tang-2022}               & \kong                                                                              & \kong                                                                                     & \kong                                                                                          & \shi                    & \kong                      & \kong                    & \shi                                                                               \\
MiniAppPerm~\cite{thung2024towards}              & \kong                                                                              & \kong                                                                                     & \kong                                                                                          & \shi                    & \kong                      & \kong                    & \shi                                                                               \\
\nameDe                                          & \shi                                                                               & \shi                                                                                      & \shi                                                                                           & \shi                    & \shi                       & \shi                     & \shi                                                                               \\ \hline
\end{tabular}
}
}
\begin{tablenotes}
            \item \footnotesize \shi means it has the corresponding feature or can achieve the requirement. 
            \item \footnotesize \kong means it does not have the corresponding feature or cannot achieve the requirement. 
        \end{tablenotes}
\end{table*}


\textbf{Existing works.} 
To address the above issues, existing works have proposed various approaches for debloating Android apps to identify and remove unused code~\cite{jiang_reddroid_2018,liu2023autodebloater,liu2024minimon,pilgun_dont_2020,xdebloat-tang-2022,thung2024towards}.
While it's not surprising to see these approaches relying on static and/or dynamic analyses to identify the to-be-debloated features, activities, or methods, we notice that they also predominantly choose to remove the target code from the APK and then resign and repackage it.
This straight-forward debloating strategy demonstrates several disadvantages (as summarized in Table~\ref{tab:comparison}):
\begin{itemize}
    \item From a security perspective, static debloating approaches with APK modification and repackaging would compromise the built-in anti-tampering mechanisms of Android~\cite{berlato_large-scale_2020,zhao2024research,merlo2021armand}, such as public key verification~\cite{signVerification} and app code integrity checks~\cite{integrity}. Accepting repackaged APKs from unknown sources also exposes end users to potential vulnerabilities or malware attacks~\cite{chen2019android,tian2017detection}.
    \item Lacks user convenience. Once a component is statically debloated from an app, it cannot be recovered without rebuilding the APK and reinstalling the app. Additionally, these approaches cannot modify the debloating schema (i.e., recover or remove components) at runtime.
    \item Android apps can consist of both DEX code and native libraries (see \mysec\ref{sec:backgroundDe}). None of the existing approaches effectively debloat both DEX and native library code together when generating a new APK.
    \item  While fine-grained debloating of DEX components has been achieved, existing approaches often cannot apply the same level of granularity in debloating the app's native code. For example, RedDroid~\cite{jiang_reddroid_2018} removes redundant native libraries designed for platforms different from the target device, but cannot remove specific native methods while preserving a library.

\end{itemize}
\vspace{0.5cm}



Furthermore, even if the user trusts the source of the repackaged APKs, the repackaging and resigning process is non-trivial, thus putting a question mark on the viability of static debloating.
In a pilot study, we attempted to repackage and resign the top 200 Google Play apps (without modification) using the latest version of tools (i.e., apktool~\cite{apktool}, zipalign~\cite{zipalign}, and apksigner~\cite{apksigner}).
The result indicated that 49 of them could not be repackaged, and another 31 repackaged APKs could not be installed or launched\footnote{The reasons which cause such observation include (but not limited to) incomplete code and/or resources, app self-checking, etc.}. 
This suggests that there exists a large proportion of commercial apps that are either reluctant or unable to be made compatible with the existing static debloating approaches.

\textbf{Our work.} 
In this paper, we introduce \nameDe, a late-stage framework for conducting \textbf{D}ynamic method-level \textbf{D}ebloating of both \textbf{D}EX and \textbf{N}ative code of An\textbf{Droid} apps without static APK modifications.
The key idea is to enable a runtime extension and hand over the task of actually eliminating unused code to the application framework of Android, thus addressing the aforementioned limitation of the existing static debloating approaches.  
\nameDe operates in two phases. 
At the offline preparation phase, it takes as input a collection of user-specified preferences with regard to what functionalities of a given subject app are not demanded, then leverages combined static/dynamic analyses on the app's APK to produce a \emph{debloating schema} which is by all means a list of methods corresponding to those unwanted functionalities.
Each method is tagged with the app's package name to prevent interference from methods with the same name in other apps.
Then, we come to the actual debloating phase of \nameDe.
Under the guidance of the schema, the runtime extension of \nameDe could then intervene in the execution of the subject app as an operating system component and prevent all types of unused code from being loaded into its memory space.

\nameDe employs a dedicated management app to configure the debloating schema of different apps, and utilizes ContentProvider~\cite{ContentProvider} to transfer each of them to the ART runtime instance of the corresponding app. 
\nameDe ensures that only the users can modify the debloating schema via the management app, while any other apps are restricted to read-only access, thus establishing a secure schema-transferring channel between the management app and the customized Android OS.
Designed in such a way, \nameDe thus requires no static APK modification of the subject apps, allowing it to be generic, of high availability, transparent to end users, and, in the meantime, compatible with the existing app security mechanisms of Android~\cite{signVerification,integrity}.

Note that Android runs the DEX and native code of an app using different mechanisms. 
Specifically, DEX methods are run via either interpreted execution, Just-In-Time (JIT) or Ahead-Of-Time (AOT) compilation (see \mysec\ref{sec:backgroundDe}), making native executable pieces for such methods difficult to be located and modified once they are produced; on the other hand, removing or altering the bytecode of such methods would violate Android's DEX file integrity verification mechanism and therefore crash the app. 
To tackle these challenges, 
\nameDe's runtime extension intercepts the DEX method invocation process of ART runtime, preventing the interpreter from processing the specified DEX methods, thereby avoiding their compilation and subsequent execution. 
Additionally, \nameDe freezes the method counter of the debloated methods to prevent them from triggering JIT or AOT compilation.

To make DEX code debloating user-friendly, \nameDe adopts a graceful termination strategy that redirects attempts of executing the debloated methods to an Activity of the management app, preventing unintended crashes while informing users of the relevant information. 
The challenge of debloating native methods, on the other hand, is to prevent loading the target methods into memory while leaving other methods unchanged.
To this end, \nameDe's runtime extension instruments both the library loading and native method invocation process of the Android OS. 
During the loading of a native library, \nameDe identifies the offsets of the to-be-debloated methods, calculates their memory addresses, and then zero-fills the body of these methods while introducing a return snippet so that subsequent invocations of these debloated methods would do nothing but go back to the call site.


To assess the performance of \nameDe, we gathered a dataset comprising 55 real-world applications and employed randomly selected debloating schemas on them.
Experiments showed that \nameDe effectively debloated all 187 DEX methods covered by the schema throughout intentionally triggered invocations, while also successfully zero-filled the code of 30 native methods during random app executions.
Further analysis revealed that \nameDe can mitigate up to 13,351 potential ROP gadgets by preventing DEX method compilation and reduce 586 ROP gadgets by debloating native methods. 
We also conducted three case studies to demonstrate \nameDe's effectiveness in mitigating the vulnerability within unused DEX and native methods.
Last but not least, after debloating by \nameDe, running the apps exhibited reduced CPU and memory usage compared to running their original versions.

\textbf{Contributions.}
To the best of our knowledge, our work is the first late-stage approach that copes with the specific challenges of conducting dynamic method-level debloating on Android apps. 
Moreover, with a special configuration, \nameDe can also be used for defending attacks via direct inter-app code
invocation (see \mysec\ref{sec:dici}).
Additionally, \nameDe may also adopt the form of eBPF-based implementation for more accessible and more friendly integration with commercial Android systems developed by third-party vendors (see \mysec\ref{sec:alterImplement}).
In summary, our approach has the following contributions:
\begin{itemize}
    \item We propose a novel dynamic debloating approach for Android apps, which conducts on-the-fly method-level debloating without static APK modifications, ensuring strong availability and transparency while being compatible with the existing Android security model.
    \item We developed a divide-and-rule strategy at the application framework level to effectively debloat both the DEX and native methods of Android apps without compromising their robustness.
    \item Our proposal reduces more than 13,000 potential ROP gadgets and mitigates known vulnerabilities in unused code of real-world apps, while also reducing their system resource consumption.
    
\end{itemize}

\textbf{Paper structure.}
The rest of the paper is structured as follows: Section \ref{sec:backgroundDe} introduces backgrounds regarding the method invocation mechanisms of the Android OS. Section \ref{sec:design} details the \nameDe design, followed by Section \ref{sec:evaluation} for its evaluation. Section \ref{sec:discussion} covers ethical issues, limitations, and an alternative implementation of our approach. \ref{sec:relatedDe} briefly compares our work with related studies and we conclude in Section \ref{sec:conclusion}.

\section{Background}
\label{sec:backgroundDe}

Android apps' code consists of DEX code and native library code. 
DEX code serves as Java-generated intermediate code, requiring additional compilation for the Android OS execution. 
In contrast, native library code is C-generated and directly executable.
The Android OS utilizes the Android Runtime (ART) to manage method invocations in both code types, establishing the execution environment and system interaction.
The Android OS maintains an independent ART instance for each app, which is initiated during the app's launch.
ART maintains an \texttt{ArtMethod} instance for each method within the app, including (1) an entry point specifying the address of the method's executable code and (2) a counter recording the invocation frequency.
DEX and native library code employ distinct mechanisms for method invocation:

\textbf{DEX method invocation.} 
Starting from Android 7.0, the ART employs a hybrid compilation mechanism for DEX methods.
Initially, the DEX methods' entry points directly to an interpreter, and their code is compiled and executed by this interpreter during runtime. 
If a method is frequently invoked, it will be compiled into native code by the Just-in-time (JIT) compiler and stored in memory cache~\cite{JIT}, with the entry point updated to the cache address.
However, if a DEX method is frequently invoked during the first few runs, it will be compiled into native code by the Just-in-time (JIT) compiler, stored in the memory cache~\cite{JIT}, and the entry point of the method is updated to the cache address. 
Meanwhile, ART generates a profile of frequently invoked methods. 
These methods will be compiled into native code by the Ahead-of-time (AOT) compiler when the device is idle and charging, and the compiled code is stored in an \textit{odex} file~\cite{AOT}. 
Subsequent runs of the app load the compiled code into memory, updating the method's entry point to the loaded memory address for direct execution without using the interpreter.


\begin{figure*}[!t]
    \centering
    \includegraphics[width=0.9\textwidth]{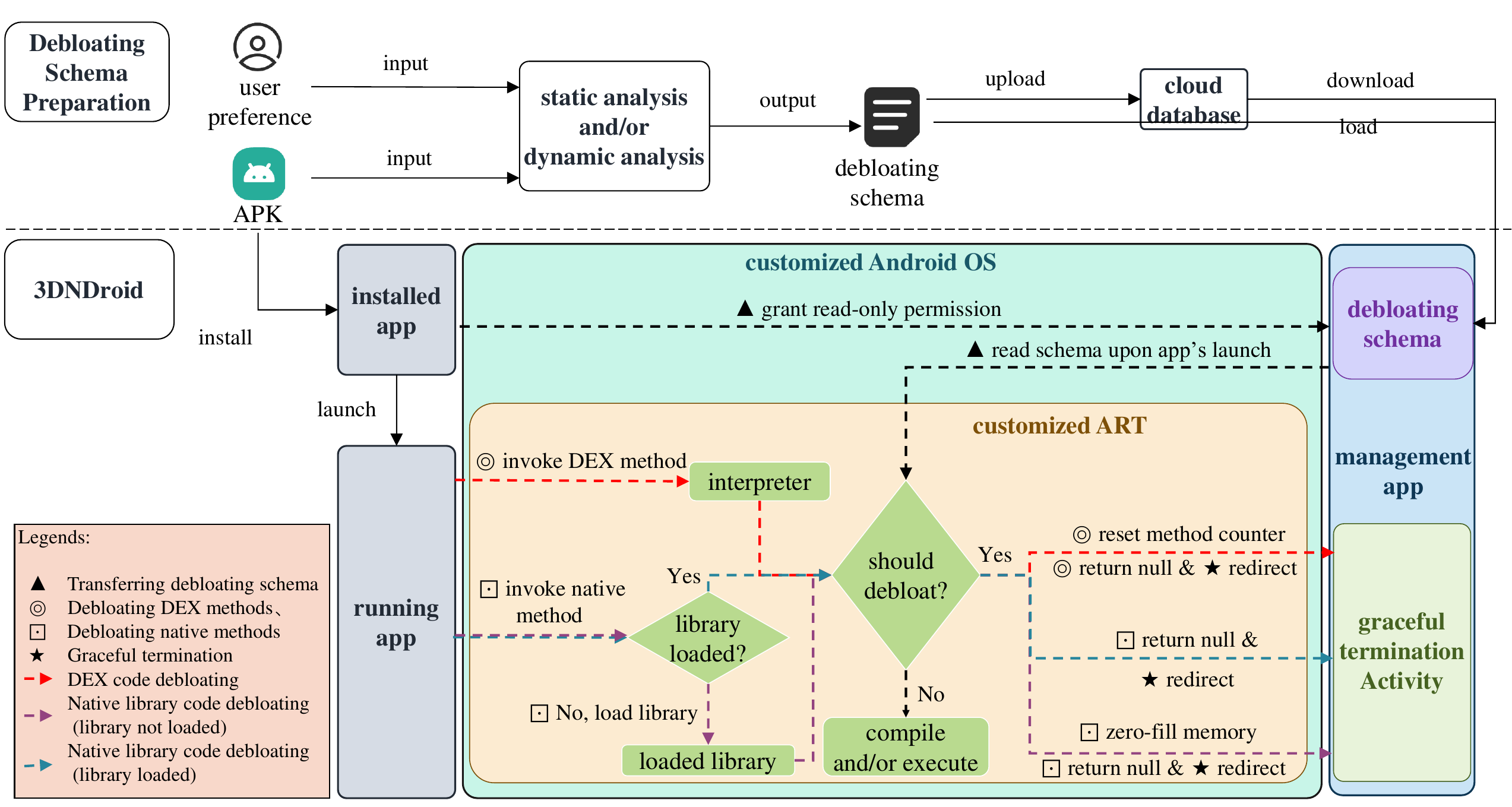}
    \caption{The overall workflow of \nameDe.}
    \label{fig:DeOverview}
    \vspace{-0.5cm}
\end{figure*}  

\textbf{Native library method invocation.}
Android allows applications to incorporate native libraries, which are \textit{so} libraries, and use JNI to invoke native methods from the Java side.
In contrast to DEX methods, native methods consist of executable instructions and do not require the interpreter. 
The initial entry point for each native library method is a null pointer.
Upon the first invocation of a native method, ART locates its memory address by searching the loaded native libraries and updates the entry point accordingly. 
If the native library has not been loaded, the Android OS searches for the library file in the installed app and loads it into the memory~\cite{ndk}.
The app can load a specific so library from the APK by using the API \texttt{System.loadLibrary()}, after which they can invoke the native method within the library.
When loading the \textit{so} library, the system reads the library's section headers, allocates memory space, and maps file sections into corresponding memory segments.
Then, ART locates the memory address of the target native method within the loaded library and updates its entry point. 
Subsequent invocations of this native method are directly executed from the updated memory address.

%

\section{Method Design}
\label{sec:design}

In this section, we first state a motivating scenario, followed by the security model and assumptions considered in the designing of \nameDe. Then, we draw an overview of the proposed approach and describe the implementation in detail.

\subsection{Motivating Scenario and Security Model}

\textbf{Motivating scenario.} Consider the scenario where an end user or an organization becomes aware that certain (rarely used) features of a widely installed app pose severe security threats due to some unpatched vulnerabilities and, therefore, seeks to temporarily block these features to reduce the exposed attack surface. 
For example, the logging feature may not be required by the end users, but the vulnerable log4j libraries affected many apps~\cite{CVE-2021-44832}, and the apps may not be patched by the developers in the short term. 
Meanwhile, the end user or an organization would also like to maintain the flexibility of reactivating the features blocked should any of them become required and well-patched. 
The whole process is like shelving the ingredients in the fridge rather than ditching them. The existing static debloating solutions would be of little value in such cases because of the above limitations.

\textbf{Security model.}
\nameDe is designed to help end users defend against unprivileged adversaries attempting to exploit vulnerabilities and ROP gadgets in unused code, and to enable users to debloat vulnerable functionalities during severe security incidents, all without modifying the APK file.
Based on this aim, we consider a security model in which the memory space of unprivileged apps is susceptible to external attacks (and therefore could use the protection of app debloating), while both the application framework and the OS kernel of Android are assumed to be intact and trusted. 
Additionally, the design of \nameDe focuses on the runtime handling of app components in the presence of a debloating scheme. 
The generation of such a schema based on an end user's preference, on the other hand, may be achieved via a number of existing approaches.
Specifically, the users may identify the to-be-debloated code using static analysis~\cite{liu2023autodebloater, pilgun_dont_2020, xdebloat-tang-2022, thung2024towards} and/or dynamic analysis~\cite{liu2024minimon, sun2016taintart, qian2014tracking}.
For native libraries, exiting works such as Jucify~\cite{samhi2022jucify} and JN-SAF~\cite{wei2018jnSAF} can help identify the unused native methods.
Preferences that could be relied on to guide such analyses would include (but are not limited to) usage~\cite{liu2024minimon}, permissions~\cite{thung2024towards}, or activities~\cite{liu2023autodebloater}, etc.
Regardless, we stress that as an early-stage preparation supporting the operation of \nameDe, code analysis is not the contribution of our work and is accordingly considered out of the scope of this paper.

\subsection{Overview}
\label{sec:DeOverview}
The overall workflow of \nameDe is illustrated in \myfig~\ref{fig:DeOverview}, which, in general, involves the cooperation between an unprivileged management app and a runtime extension module built inside the Android OS. 
Specifically, given the debloating schema of a subject app (which may be provided by practical means like a dedicated server, a cloud database, etc.), the management app of \nameDe communicates with the runtime extension module located within the ART runtime and updates the debloating configuration for the specific app upon its launching (see $\blacktriangle$).
Then at runtime, \nameDe handles the actual removal of unused components of the subject app according to a divide-and-rule strategy:
for DEX methods, it utilizes the hybrid compilation mechanism of Android,
and modifies the ART routines to intercept method invocations and prevent them from being compiled by the interpreter and the JIT/AOT compiler (see $\circledcirc$ and \mysec\ref{sec:DEXDebloat});
whereas for native library methods, it locates the memory address of to-be-debloated methods and zero-fills the corresponding memory space to remove the code during the loading of the native libraries (see $\boxdot$ and \mysec\ref{sec:nativeDebloat}).
Finally, to maintain correctness, \nameDe provides a graceful termination handling whenever a debloated method (regardless of DEX or native) is invoked (see $\bigstar$). This involves redirecting the unexpected control flows of which more details are explained in both \mysec\ref{sec:DEXDebloat} and \mysec\ref{sec:nativeDebloat}.



\subsection{Management App Design}
As described in \mysec\ref{sec:backgroundDe}, the Android OS maintains an ART instance for each app, which manages DEX method invocations conducted by the app. 
As such, the debloating schema of a specific app must be used to configure its corresponding ART instance in order to be made effective.
This thus requires an inter-procedural communication channel to facilitate the transferring of different debloating schemas obtained by the management app to the correct ART instances maintained by the OS, which needs to be able to conveniently and efficiently modify the debloating schema on-the-fly and also prevent malicious apps from evading the debloating process by altering the schema of itself or perhaps other apps. 

To establish this schema-transferring channel, we create a database in the management app to store the obtained debloating schemas and set up a ContentProvider~\cite{ContentProvider} to transfer the stored entries to the ART instances of other apps. 
We choose ContentProvider instead of the file system to conduct the schema transferring for two reasons: (1) ContentProvider provides unified interfaces for other apps to access the data provided by the management app easily; 
and (2) since Android 13 and higher versions started to employ finer granular permissions to manage the apps' storage~\cite{android13permissin}, arbitrarily granting storage access permissions will import extra security problems~\cite{alenezi2017abusing, PermissionSurvey12}.
ContentProvider allows us to define a read-only permission, allowing other apps to access the schema provided by the management app but restricting them from making modifications (see the example in Listing~\ref{code:cpPermission}).
In Android, the ART instance of an app shares the same permissions as the app itself. 
Therefore, the aforementioned permission must be granted to every third-party app so that the app's ART can have read-only access to our ContentProvider to read the debloating schema.
Understanding this necessity, we instrument the Android OS's framework and make the permission of our ContentProvider a default system permission~\cite{ContentProviderPermission}, allowing it to be automatically granted to each app during installation (without requiring static APK modifications). 
Since the granted permission is read-only and the database content is only managed by our management app, there are no potential risks to users' security and privacy.

To cope with the aforementioned management app design, we modified the routine of ART runtime, making it call our ContentProvider to read the debloating schema upon launching an app.
This ART customization also enables convenient changing of a particular app's debloating schema: by simply updating the database of the management app, an app would then be debloated according to the new schema upon restarting; 
Given that the app's APK is not statically modified and remains intact throughout the process, if the new schema withdraws the debloating claim of a certain method, that method will be immediately made effective when the app is relaunched.
This debloated code recovery applies to both DEX and native code.

\subsection{Debloating DEX Code Methods}
\label{sec:DEXDebloat}

To debloat DEX methods, \nameDe does not remove the DEX code from the app but prevents it from being compiled into native code. 
This strategy effectively mitigates the execution and code reuse risks (e.g., ROP attacks) associated with the compiled native code of such methods.
Based on the DEX method invocation mechanism introduced in \mysec\ref{sec:backgroundDe}, to safely prevent the compilation of a DEX method, two issues should be addressed:
\begin{itemize}
    \item preventing the involved DEX method from being passed to the interpreters; and
    \item ensuring that the involved method gets neither JIT-compiled nor AOT-compiled.
\end{itemize}

As introduced in \mysec\ref{sec:backgroundDe}, initially, the entry point of a DEX method directs to an interpreter.
Based on this, we instrument the ART to check if a method belongs to the debloating schema before initiating its interpretation.
Here, we leverage a hash set to store the debloating schema to improve running performance.
The instrumented ART can obtain the package name of the current running app. 
While inspecting each method, the instrumented ART verifies whether the corresponding package name matches the current running app, ensuring that debloating is applied to the correct app.
Upon seeing a ``red flag'', the instrumented ART intercepts the method invocation and bypasses the interpreter by returning null as if the method body is empty, thus preventing further compilation processes and returning control flow to the method caller.
Meanwhile, given a so-debloated method, \nameDe resets the invocation counter within its \texttt{ArtMethod} instance (see \mysec\ref{sec:backgroundDe}) to avoid triggering the JIT/AOT compilation that would otherwise write its native executable snippet into the app's \textit{odex} file.
Note that although the native code is not generated, the entry point of a debloated method remains unchanged, meaning that subsequent invocations of the same method would still be directed to the interpreter and subject to debloating checks.
This allows restoring a debloated method by just removing it from the schema and, therefore, letting it pass the debloating check.
During debloating, \nameDe filters out all the standard Java or Google APIs, ensuring that all the debloated methods are from the target app's own code. 
Through the above process, \nameDe effectively debloates DEX methods without having to statically rewrite the DEX files.

\textbf{Graceful Termination.}
Although the debloated methods are rarely used, to ensure a user-friendly experience in cases users accidentally trigger these methods, we integrated a graceful termination feature into \nameDe so that once a debloated method is invoked, the management app will be informed and present the information to users (for example, by popping an AlertDialog which tells the user that a certain method is not executed due to debloating).
For DEX methods, specifically, once a method invocation is intercepted according to the schema, the instrumented ART launches the management app's informing Activity before passing the debloated method to the interpreter and letting the latter return with null.
Note that this feature is optional, i.e., \nameDe can conduct the debloating process without this graceful termination.

\textbf{Recovery.} To recover the debloated DEX methods, users just need to remove them from the debloating schema and relaunch the app. The customized ART will read the new schema upon the app's launch, and therefore the invocation of certain DEX methods will no longer be intercepted.


\subsection{Debloating Native Library Methods}
\label{sec:nativeDebloat}
In addition to DEX code, native libraries are extensively used by app developers for their capability to conduct low-level operations (e.g., accessing hardware or performing I/O operations)~\cite{ruggia2023dark} or carry out CPU-intensive tasks (e.g., image processing and video encoding)~\cite{hoepman_nativeprotector_2016,almanee2021too,sun2014nativeguard}.
As described in \mysec\ref{sec:backgroundDe}, unlike DEX methods, executing native library methods does not require interpretation or compilation. 
Simply intercepting invocations of native library methods is thus insufficient, as the native method code is already loaded into memory and could be exploited as gadgets to facilitate code reuse attacks.
In addition, upon the first invocation of a native method, ART locates its memory address (see \mysec\ref{sec:backgroundDe}).
However, preventing ART from obtaining the address of the native method would lead to app crashes, as the entry point of the native method is initially a null pointer.
Moreover, when the Android OS loads the native library, the starting offset of each section must align with the memory page size. 
Hence, it is impossible to separately load each native method from the library when multiple native methods exist within the same memory page. 
Last but not least, we also need to refrain from modifying the kernel or altering system calls for library loading; otherwise, it could introduce additional security risks~\cite{hei2013two}.


To address the issues above and achieve native method debloating without having to bring its modifications all the way down to the Linux kernel, \nameDe strategically locates the to-be-deloated methods in a native library and removes their code body from memory on-the-fly through zero-filling.
Specifically, while loading a native library containing methods claimed by the debloating schema, \nameDe captures the starting and ending offsets of the target method from the library file.
This is achieved by parsing the headers of the \textit{so} library, reading and storing two additional headers called \texttt{dynsym} and \texttt{dynstr}, which record the offsets of all the functions in the library.
Similar to the debloating process for DEX code, \nameDe verifies the package name of native methods to ensure it matches the currently running app.
Following the native library loading process described in \mysec\ref{sec:backgroundDe}, after the system allocates the memory space of the native library, \nameDe computes the method's memory space by adding the two offsets to the allocated memory address.
Then, after the system maps the library file into the memory, \nameDe locates and zero-fills the memory space of the target methods to erase their code from the loaded content.

Since the native method debloating is conducted during the library loading process, and each library is only loaded once, while the app is loading a native library, \nameDe checks if any method belongs to the debloating schema and debloats all the claimed native methods from the library.

\begin{figure}[!t]
    \centering
    \includegraphics[width=0.48\textwidth]{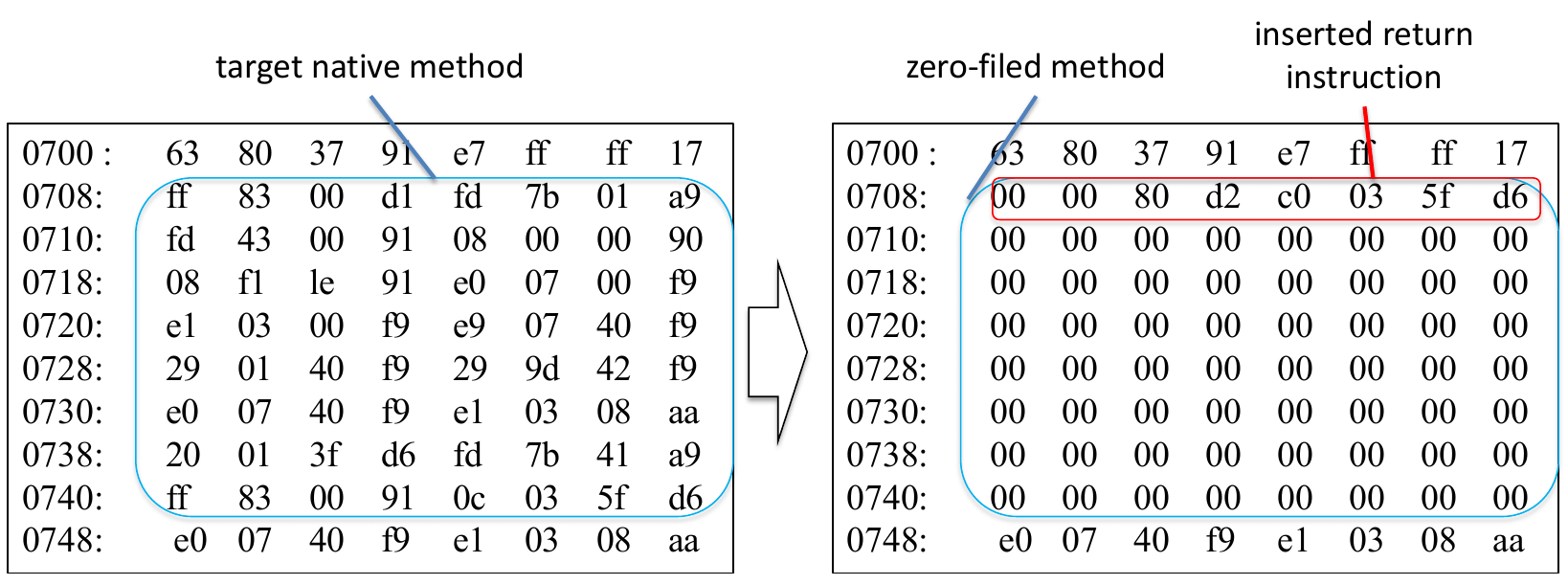}
    \caption{An example of zero-filling the debloated native method and inserting return instruction.}
    \label{fig:zero-fill}
    \vspace{-0.5cm}
\end{figure}

Additionally,  \nameDe inserts a return instruction to redirect the control flow back to the call site when a debloated native method is invoked.
\myfig~\ref{fig:zero-fill} demonstrates an example of this process.
Since every native method inherently includes a return instruction, the minimum size of a native method is eight bytes. This guarantees that the inserted return instruction remains within the code space of the debloated native method, avoiding any impact on other code segments.
With these processes in place, when a debloated native method is invoked, ART can still fetch its memory address and update its entry point accordingly although its code has been removed and replaced.
Subsequent invocations of the method will be directed to the return instruction without causing unintended crashes, hence preserving the robustness of the debloated app.

\textbf{Graceful Termination.}
Again, the debloating of native methods is also included in the graceful termination handling of \nameDe. Except for this time, the instrumented ART launches the management app's informing Activity when locating the entry point of a debloated native method during the JNI routine.
Moreover, the graceful termination is conducted independently from the return instruction inserted in the zero-filled space, i.e., when handling such a debloated invocation, the return snippet replacing the callee's code body acts on its own in the background to redirect control flow back to the caller, while the informing Activity is launched by ART to display essential information in the foreground.

\textbf{Recovery.} 
Like DEX code debloating, users can remove debloated methods from the schema and relaunch the app to restore the native library methods. Once the new schema is applied, the app will load the corresponding library normally without zero-filling those methods.

\section{Evaluation}
\label{sec:evaluation}
In this section, we aim to comprehensively evaluate \nameDe by answering these research questions:
\begin{description}
    \item [RQ1:] Can \nameDe effectively debloat DEX and native methods in Android apps?

    
    \item [RQ2:] How many potential code reuse gadgets can be reduced after \nameDe's debloating? 
    
    \item [RQ3:] Can \nameDe help mitigate vulnerabilities in apps?

    \item [RQ4:] What are the resource consumption differences of running applications with and without \nameDe's debloating?
\end{description}

\subsection{Experiment Setup and Data Collection}
\label{sec:experimentSetup}
To answer these RQs, we implemented a prototype of \nameDe with AOSP 13 on the Pixel 7 Pro with 8GB RAM and 256GB storage. 
All the experiments are conducted on this device, and we connected it to our workstation to collect corresponding experiment results, e.g., logs and dump files. 
To assess the performance of \nameDe in RQ4, we flashed the original AOSP 13 image without \nameDe to the same device for comparison in \mysec\ref{sec:ResEvaluate}.

\textbf{Schema collection.} 
To address RQ1, 2, and 4, we tested 55 real-world applications collected from Google Play. 
These apps are randomly selected from the top 40 apps in each category listed in AndroidRank~\cite{androidRank}.
Since \nameDe requires a debloating schema as input (i.e., list of methods to be debloated), we leverage the state-of-the-art static debloating tool, AutoDebloater~\cite{liu2023autodebloater}, to generate the schema for each app.
During the app collection process, any apps that failed to generate call graphs were excluded, as we cannot obtain the corresponding schema of these apps.



Among the 55 apps, 50 were used to evaluate \nameDe's effectiveness in DEX method debloating (see \mysec\ref{sec:DEXEvaluate}), comprising 45 non-commercial apps and five commercial apps.
For the non-commercial apps, we randomly selected one Activity from each app and leveraged AutoDebloater to conduct forward and backward slicing on the app's call graph and obtain the related methods as the debloating schema. 
We avoided selecting Activities like \texttt{MainActivity} and \texttt{SplashActivity} when generating the schema, as these Activities serve as app entry points, and debloating them would prevent the app from launching.
If AutoDebloater failed to generate the schema for one app, we selected another app and repeated the process until we collected 45 apps, each with a non-empty schema. 
The debloating schema for each app encompassed a variable number of methods, ranging from 1 to 168, with an average of 24 methods.

The other five apps are commercial apps, including Adobe Reader~\cite{adobe}, Airasia~\cite{airasia}, Discord~\cite{discord}, Homeworkout~\cite{homeworkout}, and Duolingo~\cite{duolingo}. 
These commercial apps cannot be statically modified by third parties due to anti-tampering mechanisms~\cite{berlato_large-scale_2020,zhao2024research,merlo2021armand} and thus could only be debloated using \nameDe.


\begin{figure*}[!t]
    \centering
    \includegraphics[width=\textwidth]{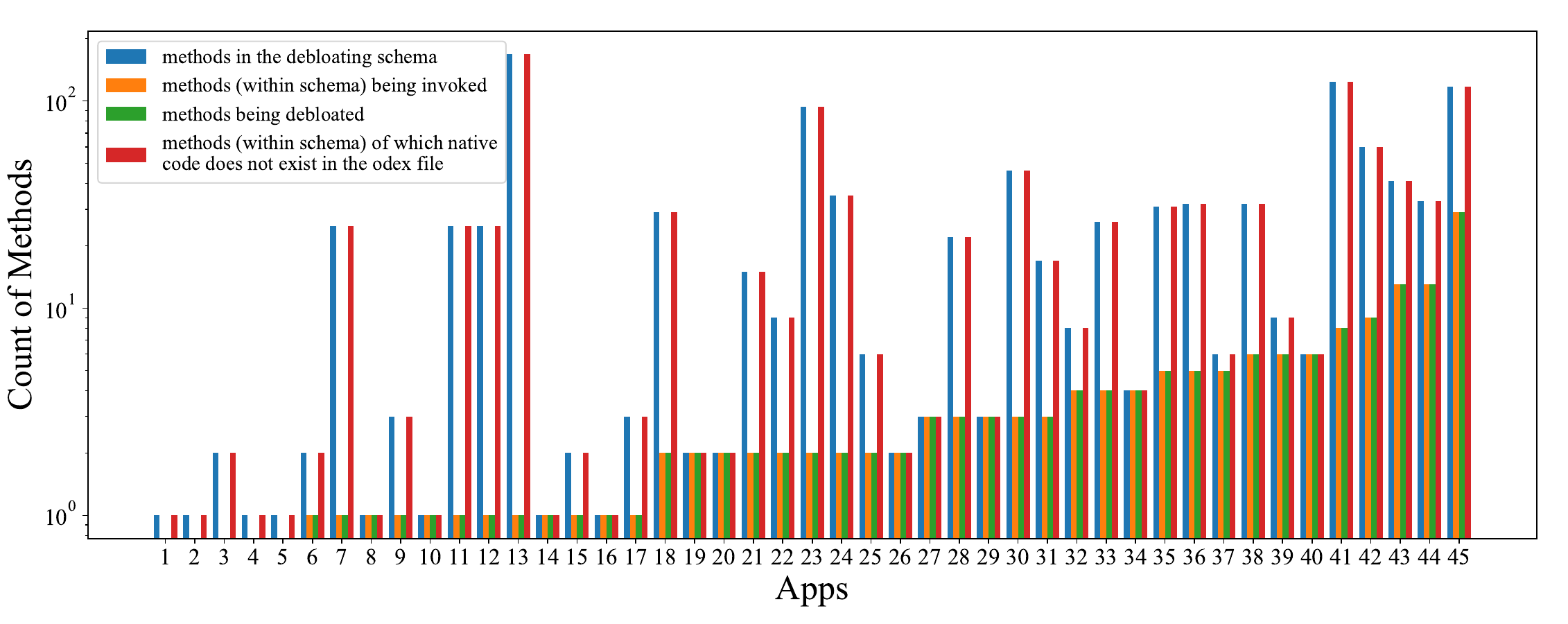}
    \caption{The counts of various statistics for each app, encompassing the number of methods in the debloated schema, the number of invoked methods covered by the schema, the methods debloated by \nameDe, and the methods of which native code does not exist in the corresponding \textit{odex} file.}
    \label{fig:debloat_invoke}
    \vspace{-0.5cm}
\end{figure*}

To evaluate \nameDe's effectiveness on debloating native methods, we also collected apps containing native libraries from the top 40 apps in each category and randomly selected six apps from them (see \mysec\ref{sec:NativeEvaluate}). 
Remarkably, one of these apps (\path{kha.prog.mikrotik}) is also included in the 50 apps used for DEX method debloating evaluation. 
This app demonstrates \nameDe's capability of debloating both native and DEX methods concurrently.
Different from collecting the DEX method schema, AutoDebloater is not designed for generating native code schema.
Therefore, we employed Jucify~\cite{samhi2022jucify}, the state-of-the-art tool that generates call graphs combining DEX code and native code of Android apps, to create the call graphs for these six apps. 
Subsequently, for each app, we randomly chose five native methods from the generated call graph to serve as the debloating schema.

\subsection{Effectiveness on Debloating Methods}
\subsubsection{Debloating DEX Methods}
\label{sec:DEXEvaluate}
As \nameDe is a dynamic debloating tool that operates without modifying the APK file, metrics commonly used in previous works~\cite{jiang_reddroid_2018,xdebloat-tang-2022}, like reduced APK sizes and removed lines of code, are not applicable for \nameDe's evaluation.
To evaluate \nameDe's effectiveness in debloating DEX methods, we run the apps and dynamically monitor the invocation and debloating of methods listed in the debloating schema.
Specifically, we use Monkey~\cite{monkey}, the official Android automatic testing tool, to automatically run the apps with their debloating schemas as input. 
For each of the 45 non-commercial apps collected in \mysec\ref{sec:experimentSetup}, Monkey randomly generates 3,600 events (e.g., touches, swipes, or Activity launches), with a one-second delay between every two events.
20\% of the events involve initiating Activities, ensuring that Monkey covers a diverse range of Activities in the tested app.
Each app requires approximately one hour for automated execution. 
To verify \nameDe's ability to debloat invoked methods within the schema, we modify the AOSP to log all invoked methods and debloated methods in the logcat~\cite{logcat}, the official Android logging tool.

As introduced in \mysec\ref{sec:backgroundDe}, the frequently invoked methods will be recorded into a profile and subsequently compiled into native code and stored in the \textit{odex} file.
However, the AOSP developer documentation~\cite{AOT} does not explicitly specify the minimum times of invocations for a method to be recorded in a profile.
To address this problem, we ran two apps for an hour without any methods debloated by \nameDe and checked the invoked methods as well as the compiled \textit{odex} file. 
The result showed that, after an hour of running, the compiled native code of invoked methods appeared in the app's \textit{odex} file.  
Based on this observation, to further verify that the debloated methods are not recorded in the profile, after running each app for an hour using Monkey, we trigger the profile-based AOT compilation (the related command is presented as Listing~\ref{code:AOTcompilation})
and verified the absence of native code for the debloated methods in the compiled \textit{odex} file.




\textbf{Experiment results.}
Since Monkey generates random events to run the apps, it cannot guarantee that all the debloated methods are invoked during runtime. 
However, obtaining the operation trace for triggering each debloated method is non-trivial.
Therefore, we mainly consider the methods that are invoked during the automatic running.

\textit{Non-commercial apps.}
\myfig~\ref{fig:debloat_invoke} illustrates the count of different statistics for the 45 non-commercial apps, sorted based on the number of invoked methods within the schema of each app.
Specifically, for each app, we analyzed the methods within the schema (blue bars) and assessed the invoked methods included in the schema during testing (orange bars) by checking the logcat.
We also analyzed the logcat to obtain the methods debloated by \nameDe (green bars).
Finally, we examined each app's \textit{odex} file to confirm the absence of native code for every debloated method, which means that these methods are not compiled (red bars).

Due to the limitation that Monkey explores only a subset of methods within the schema for each app, there is a notable gap between the debloating schema and the invoked methods.
Therefore, we employ logarithmic scaling to display the method counts, which helps mitigate the perceived quantitative gap while preserving the relative size relationships.
In total, the debloated schemas for the 45 apps encompass 1,077 methods, whereas only 162 methods are invoked. 
Five apps have no invoked methods within the schema and, consequently, exhibit no debloating records in the logcat.
For the remaining 40 apps, the lengths of the orange and green bars are identical, indicating that \nameDe successfully debloated all invoked methods within the schema. 
Moreover, each app's blue and red bars have matching lengths, illustrating that none of the methods within the debloating schema were compiled into native code.


\textit{Commercial apps.}
Regarding the five commercial apps collected in \mysec\ref{sec:experimentSetup}, their larger size and the requirement for login make it difficult for Monkey to explore these apps automatically.
To overcome this limitation, we manually logined and ran each app for 20 minutes without debloating and recorded all the invoked methods. 
After that, we randomly selected five invoked methods for each app as the debloating schema.
We reinstalled each app, applied the debloating schema, and manually re-ran the app to check if those methods were debloated.
The results demonstrate that all 25 DEX methods within the schema were successfully debloated, with invocations intercepted and redirected to graceful termination by \nameDe (as described in \mysec\ref{sec:DEXDebloat}).

The above results demonstrate that \nameDe effectively debloats DEX methods in real-world apps, intercepting all invocations of debloated methods and preventing them from being compiled into native code.



\subsubsection{Debloating Native Methods}
\label{sec:NativeEvaluate}
Similar to the DEX code debloating evaluation, we tested the six apps containing native libraries using Monkey on \nameDe with their corresponding schema as input.
During the app running, our modified AOSP recorded all the loaded native libraries and invoked native methods in the logcat.
Unlike DEX code, native code is directly loaded into memory without compilation, simplifying the verification for the absence of native methods' code. 
When a debloated native method is invoked, our customized Android OS prints out the memory content of the allocated method address after loading the corresponding library. 
This logging process verifies whether the code for the debloated native method is zero-filled, as explained in \mysec\ref{sec:nativeDebloat}.

As described in \mysec\ref{sec:nativeDebloat}, while the Android OS is loading a native library, all methods claimed by the debloating schema within this library are debloated by \nameDe. 
To initiate the library loading process for the six apps, we utilize Monkey to run the apps automatically on \nameDe with 2,000 randomly generated events. 
For each app, we input five native library methods collected in \mysec\ref{sec:experimentSetup} as the debloating schema. 
While running the apps, we recorded the logcat results for each app and inspected the debloated methods.

\textbf{Experiment results.}
For each of the six tested apps, all five native methods were successfully debloated, indicating that \nameDe successfully located the memory addresses of these native methods and zero-filled their corresponding native code in memory.
However, one app (\path{com.nikon.snapbridge.cmru}) experienced a crash during testing, caused by debloating a native method crucial for the app's regular operation. 
Since our debloating schema for native methods was randomly selected, this issue could be mitigated by obtaining a more accurate schema that excludes these crucial methods.
Despite the crash experienced by this app, the logcat result indicates that all five methods included in its debloating schema were successfully debloated.
Moreover, \nameDe successfully debloated native methods of the app \path{kha.prog.mikrotik}, which is also employed to assess \nameDe's debloating ability of DEX code. 
This result demonstrates \nameDe's capability to debloat DEX and native methods concurrently.

\begin{center}
    \vspace{-2ex}
    \begin{mybox}[boxsep=0pt,
	boxrule=1pt,
	left=4pt,
	right=4pt,
	top=4pt,
	bottom=4pt,
	]

    \textbf{Answer for RQ1: }
        \nameDe effectively debloats randomly selected 187 DEX methods and 30 native methods in 55 real-world apps.
        Moreover, it can concurrently debloat both DEX and native methods in the same app.
         
    \end{mybox}
    \vspace{-1ex}
\end{center}

\begin{figure}[!t]
    \centering
    \includegraphics[width=0.4\textwidth]{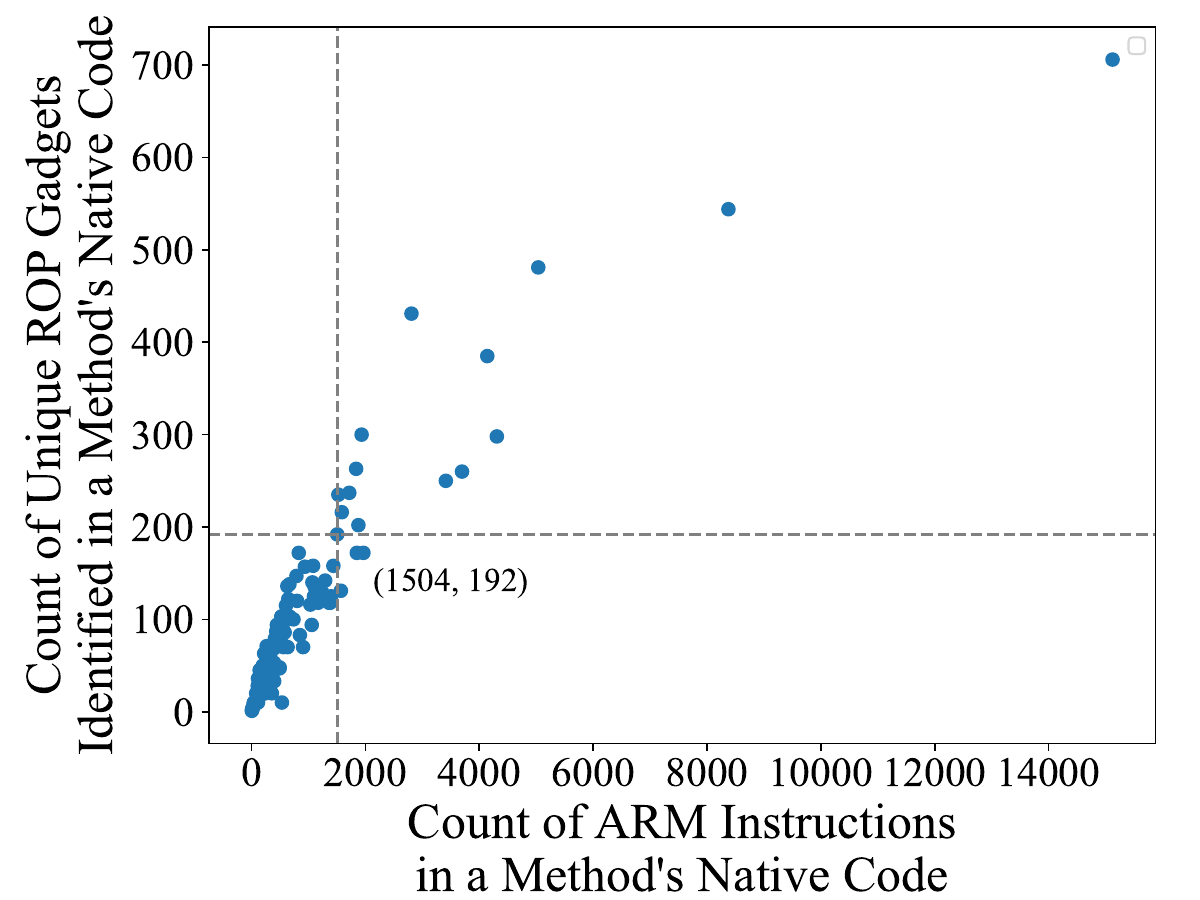}
    \caption{The number of ROP gadgets found in various sizes of DEX methods compiled in native code. The bottom-left portion of the intersection of the two dashed lines contains 90\% of the methods.}
    \label{fig:methodSizeGadgets}
    \vspace{-0.5cm}
\end{figure}

\subsection{Evaluation on Reducing ROP Gadgets}
\label{sec:ROPEvaluate}

One of the primary objectives of this paper is to reduce the attack surface for code reuse attacks within Android apps, particularly focusing on mitigating Return-Oriented Programming (ROP) attacks. 
In ROP attacks, an attacker manipulates the call stack to hijack program control flow and then executes carefully chosen machine instruction sequences that are already present in the memory, known as ``gadgets''.
Each gadget consists of ARM instructions that load or store specific values into registers or jump to specific addresses for execution.
For example, the ARM instruction \texttt{ldr x30, [sp, \#0x18]} loads data from the memory location indicated by the stack pointer plus \texttt{0x18} into the \texttt{x30} register.
Following the works~\cite{raja2017return,sun2016blender}, which emphasized that the quantity and variety of ROP gadgets are direct measures for the feasibility of ROP attacks, our evaluation focused on these metrics.
By debloating the DEX and native methods, the corresponding compiled code (ARM instructions) is removed from memory, thereby reducing the available ROP gadgets.
To comprehensively assess the effectiveness of \nameDe in reducing ROP gadgets, we conducted separate evaluations on DEX and native code.

\textbf{Searching for ROP gadgets in compiled DEX methods.}
For the 45 non-commercial apps, since only the ARM instructions can be utilized as ROP gadgets, we compiled all the DEX code of these apps into native code (i.e., ARM instructions) using \textit{dex2oat}~\cite{dex2oat}.
This process generates an \textit{odex} file for each target app, which stores all its DEX code and the corresponding compiled native code~\cite{AOT}.
After that, we utilized \textit{oatdump}~\cite{oatdump} to parse the \textit{odex} file and extracted the file offset of each method stored in the \textit{odex} file.
To evaluate the reduction of ROP gadgets after \nameDe debloats the DEX methods, we leveraged the debloated methods of the 45 apps (162 in total) obtained in \mysec\ref{sec:DEXEvaluate} as the target and located their corresponding offsets in the \textit{odex} files.


With the above preparation, we collected the debloated methods' ROP gadgets from the compiled \textit{odex} file. 
We utilized a commonly used tool called \textit{ROPgadgets}~\cite{ROPgadget}, which employs Galileo algorithm~\cite{shacham2007geometry} to search ROP gadgets and outputs all the unique ROP gadgets found in the input binary file.
When searching for gadgets in the \textit{odex} file, we specified a search range consisting of each method's native code starting and ending offset.
This specification enabled \textit{ROPgadgets} to search for ROP gadgets only within the native code of each method.


\begin{figure*}[t!]
  \centering
  \begin{subfigure}[b]{0.4\textwidth}
    \includegraphics[width=0.9\textwidth]{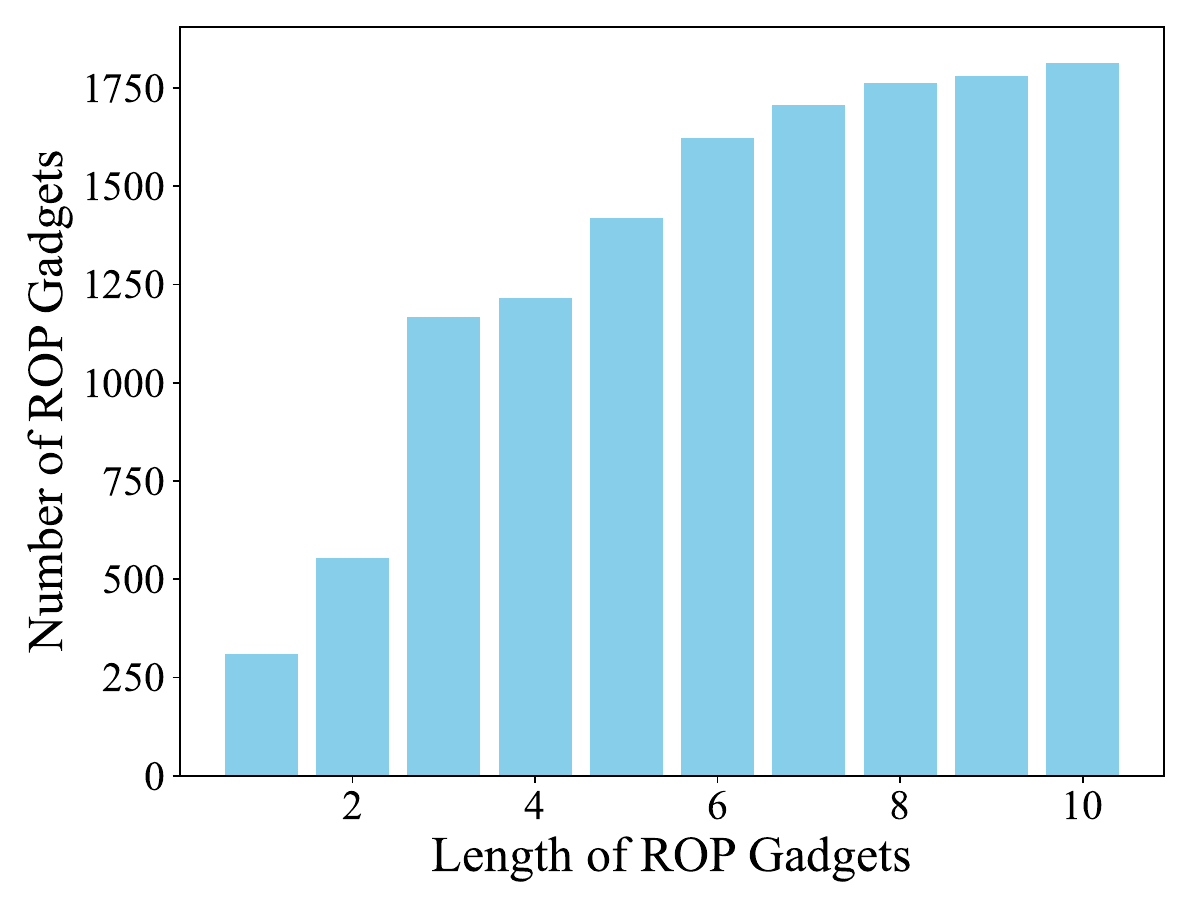}
    \caption{The count of ROP gadgets with various\newline lengths.}
    \label{fig:gadgetsLength}
  \end{subfigure}
  \begin{subfigure}[b]{0.4\textwidth}
    \includegraphics[width=0.9\textwidth]{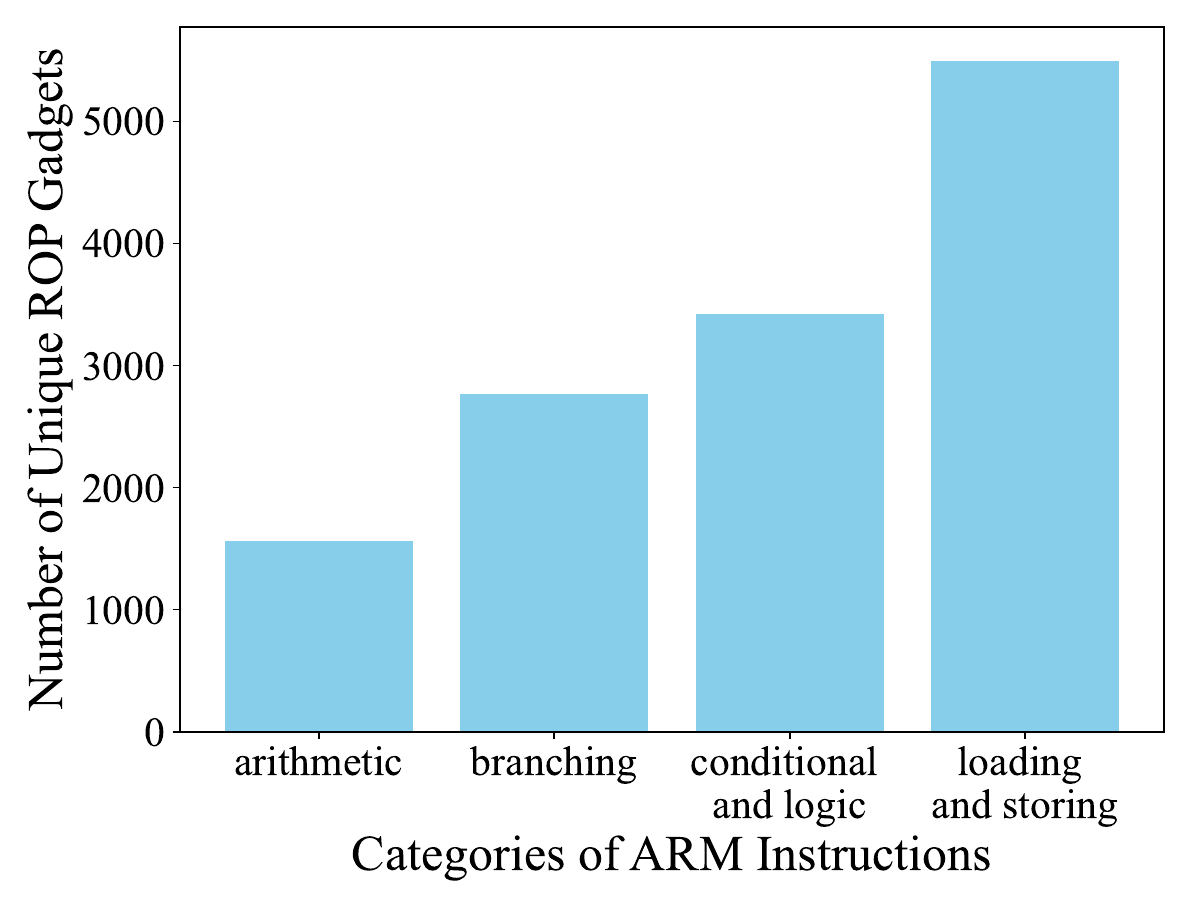}
    \caption{\centering The count of ROP gadgets started with various types of ARM instructions.}
    \label{fig:GadgetsCategory}
  \end{subfigure}
  \caption{The statistics of ROP gadgets found in debloated DEX methods post-compilation.}\label{fig_rq3}
  \vspace{-0.5cm}
\end{figure*}

\textbf{Experiment results for compiled DEX methods.}
Overall, we identified 13,351 ROP gadgets in the debloated 162 methods, with an average of 82.4 gadgets per method.
\myfig~\ref{fig:methodSizeGadgets} illustrates the distribution of gadget numbers found in compiled DEX methods of different sizes, measured by the number of ARM instructions.
We observed that 90\% of the methods consisted of less than 1,504 instructions but could contain up to 192 unique ROP gadgets. 
This highlights the significance of debloating DEX methods, as even a small method can harbor a substantial number of ROP gadgets.

To better understand the severity of these ROP gadgets, we analyze the number of gadgets of varying lengths and categorize them based on the first ARM instruction for each gadget.
Firstly, the distribution of gadgets with various lengths is illustrated in \myfig~\ref{fig:gadgetsLength}.
We observed that more than half of the collected ROP gadgets (i.e., 7,062) contained at least seven ARM instructions. 
These gadgets consist of sequential ARM instructions, which may facilitate attackers in crafting sophisticated attack payloads~\cite{raja2017return,cheng2014ropecker,hanna2013juxtapp}.

Secondly, aligning with the documentation from ARM developers~\cite{arm64document} and prior research~\cite{raja2017return,cheng2014ropecker}, we classified the ROP gadgets into four distinct categories based on their first ARM instruction.
The distribution of gadgets with various lengths is illustrated in \myfig~\ref{fig:gadgetsLength}.
Specifically, the four categories include arithmetic instructions (performing various operations on data in registers), branching instructions (changing the execution flow of a program), conditional and logic instructions (comparing the values in the registers and executing conditional instructions based on the comparison result), and loading and storing instructions (loading data from memory to registers or storing data from registers back into memory).
In the results, we observed that the predominant category of gadgets starts with loading and storing instructions, constituting more than 41\% of all gadgets. 
These gadgets offered attackers a variety of options for loading and storing data between registers and memory.
In summary, \nameDe's debloating of DEX methods significantly reduces a variety of potential ROP gadgets.





\textbf{Searching ROP gadgets in native methods.}
The native library methods of an app are stored in the \textit{so} library, an ELF format file~\cite{lu1995elf}. 
Since these native methods consist of ARM instructions, we can directly employ \textit{ROPgadgets} to search for ROP gadgets within the target native libraries. 
Similar to how we conducted native methods debloating in \mysec\ref{sec:nativeDebloat}, we obtained the file offset of each debloated native method as well as their sizes from the ELF headers of the \textit{so} library.
In each of the six apps tested in \mysec\ref{sec:NativeEvaluate}, the five debloated native methods were analyzed using \textit{ROPgadgets}. 
By specifying the starting and ending offset (i.e., starting offset plus size) of each debloated method, \textit{ROPgadgets} output all the corresponding ROP gadgets within the \textit{so} file.

\textbf{Experiment results for native methods.}
The average size of the 30 debloated native methods is approximately 265.6 ARM instructions. 
From these methods, we gathered a total of 586 ROP gadgets, averaging 19.5 gadgets per method. 
Similar to the gadgets found in compiled DEX methods, over half of these gadgets (298) comprise at least six ARM instructions. 
The predominant category of gadgets (238) also starts with loading and storing instructions, accounting for about 40.6\% of all gadgets. 
These findings demonstrate \nameDe's capability to significantly reduce potential ROP gadgets after debloating both DEX and native methods in Android apps.


\begin{center}
    \vspace{-2ex}
    \begin{mybox}[boxsep=0pt,
	boxrule=1pt,
	left=4pt,
	right=4pt,
	top=4pt,
	bottom=4pt,
	]

    \textbf{Answer for RQ2:}
        \nameDe reduces 13,351 potential ROP gadgets in compiled DEX methods and 586 gadgets in native code by debloating both DEX and native methods. This substantial reduction significantly decreases the attack surface for ROP attacks.
    \end{mybox}
    \vspace{-1ex}
\end{center}

\subsection{Vulnerability Mitigation}
\label{sec:vulnerability}
As described in \mysec\ref{sec:intro}, another primary objective of \nameDe is to mitigate vulnerabilities in Android apps. 
In real-world scenarios, users can temporarily debloat these vulnerable methods using \nameDe if they do not require the functionality or when they find the specific app components encounter severe security incidents and wait until developers release a patch to fix the vulnerability.
By debloating the specific vulnerable methods, they could continue to use other functionalities of that app and mitigate the vulnerability.

To evaluate \nameDe's efficacy in achieving this goal, we conducted case studies using known CVEs from the public CVE database~\cite{CVEProgram}.
We selected three DEX code vulnerabilities and one native library vulnerability, leveraging datasets from \textit{PHunter}~\cite{xie2023precise} and \textit{LibRARIAN}~\cite{almanee2021too}, which provide information on CVEs and affected apps related to DEX code and native libraries, respectively.
For each CVE, we identified the corresponding vulnerable methods by manually reviewing the CVE descriptions and generating the debloating schema. 
Subsequently, we applied the schema to \nameDe to debloat these vulnerable methods in the app affected by each CVE. 
After that, upon the app's launch, we leverage the ART to directly invoke each of these methods of the target apps via Java reflection and check if these methods are debloated by \nameDe.

\textbf{Case 1: CVE-2019-20444~\cite{CVE-2019-20444} (DEX code).} 
Within versions of the \path{Netty} library~\cite{netty} prior to 4.1.44, the method named \texttt{splitHeader()} of the class \texttt{HttpObjectDecoder} does not check whether an HTTP header lacks a colon, and the incorrect header might be interpreted as a separate header with an incorrect syntax or an invalid fold. 
This vulnerability allows attackers to conduct HTTP smuggling attacks~\cite{jabiyev2021t}.
The app \path{com.btcontract.wallet} (version 2.4.27) contains this vulnerable version of \path{netty} library as well as the vulnerable method \texttt{splitHeader()}.
If users do not require any HTTP-related features of the app (e.g., browsing websites, etc.), they can debloat this method from the app \path{com.btcontract.wallet} using \nameDe; therefore, the attacker cannot leverage this method to launch HTTP smuggling attacks.

\textbf{Case 2: CVE-2020-26939~\cite{CVE-2020-26939} (DEX code).}
This CVE relates to the OAEP Decoder in the Bouncy Castle library~\cite{BCJava} prior to version 1.61. 
Sending invalid ciphertext that decrypts to a shorter payload in the OAEP decoder may trigger an early exception, potentially exposing RSA private key details.
This vulnerability is specifically associated with the method named \path{decodeBlock()} of the class \path{OAEPEncoding}.
After debloating this method from the affected app \path{com.xabber.android} (version 2.6.6.644) using \nameDe, the method becomes inexecutable, preventing the potential private key exposure.


\textbf{Case 3: CVE-2024-32876~\cite{CVE-2024-32876} (DEX code).}
\path{org.schabi.newpipe} is a third-party client app for streaming YouTube videos~\cite{NewPipe}. It supports exporting and importing backups of user profiles. However, in versions 0.13.4 through 0.26.1, because no validation is performed on the imported files, it had a vulnerability that allowed Arbitrary Code Execution if a malicious backup file was imported. 
The vulnerability is related to the method \path{importDatabase()} of class \path{BackupRestoreSettingsFragment}. 
Since loading the backup file is not a main feature of this app, users can leverage \nameDe to temporarily debloat this method until the app developers release patches, without affecting streaming YouTube videos.

\textbf{Case 4: CVE-2014-0191~\cite{CVE-2014-0191} (Native library code)}. This CVE is related to the native method \texttt{xmlParserHandlePEReference} in the native library \texttt{libxml2.so}~\cite{libxml2} before version 2.9.2. 
This vulnerability involves loading external parameter entities regardless of entity substitution or validation settings. 
This oversight enables remote attackers to launch denial-of-service (DOS) attacks (resource consumption) via a crafted XML document.
The application \texttt{com.amazon.kindle} (version 8.29.0.100) is vulnerable to this exploit. 
After debloating this native method within the native library using \nameDe, adversaries cannot execute the method via a crafted XML document, preventing excessive consumption of system resources that leads to DOS attacks.
\begin{figure*}[ht]
  \centering
  \begin{subfigure}[b]{0.3\textwidth}
    \includegraphics[width=1.0\textwidth]{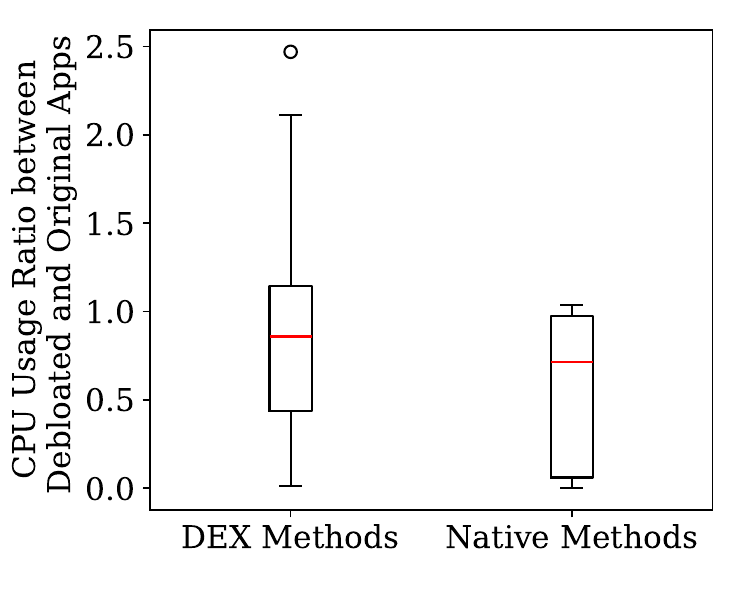}
    \vspace{-0.5cm}
    \caption{CPU usage ratio}\label{cpu}
  \end{subfigure}
  \begin{subfigure}[b]{0.3\textwidth}
    \includegraphics[width=1.0\textwidth]{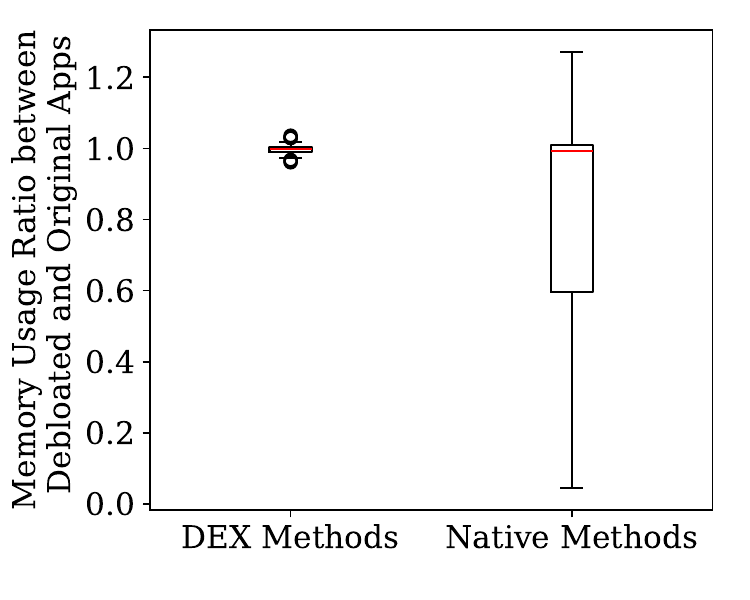}
    \vspace{-0.5cm}
    \caption{Memory usage ratio}\label{memory}
  \end{subfigure}
  \caption{The ratio of the resources (i.e., CPU and memory) usage difference between debloated apps and original apps. The red bars represents the median values.}
  \label{fig_rq4}
  \vspace{-1em}
\end{figure*}

\begin{center}
    \vspace{-2ex}
    \begin{mybox}[boxsep=0pt,
	boxrule=1pt,
	left=4pt,
	right=4pt,
	top=4pt,
	bottom=4pt,
	]

    \textbf{Answer for RQ3:}
        The four case studies illustrate that \nameDe can effectively mitigate vulnerabilities in Android apps by debloating the vulnerable methods. Users can leverage \nameDe to block the undesired vulnerable methods until the patch is released.
        
    \end{mybox}
\end{center}


\subsection{Performance Evaluation}
\label{sec:ResEvaluate}

Following Tang et al.\cite{xdebloat-tang-2022}'s work, we estimated the resource overhead introduced by \nameDe.
Specifically, we compared the resource consumption of \nameDe against the original AOSP system while running apps.
To do so, we utilized Monkey to automatically run the 45 non-commercial apps collected in \mysec\ref{sec:experimentSetup}.
While running the apps with Monkey, we recorded their event traces and CPU and memory usage using the \textit{adb} shell command every minute.
However, measuring resource usage can encounter various unexpected scenarios, such as an unstable network. 
Therefore, we needed to remove outlier data points.
To achieve this, we followed prior studies~\cite{jiang_jred_2016, carter2009a, schwertman2004simple} and employed the boxplots to remove the outlier data points.
Specifically, in a boxplot, the top and bottom of a box represent the third (Q3) and first (Q1) quartiles of a data group, respectively, and the Interquartile Range (IQR) is defined as $Q3-Q1$.
The maximum and minimum values are defined as $Q3+1.5\times IQR$ and $Q1-1.5\times IQR$ in a boxplot.
Data points outside of these maximum and minimum values were considered outliers and removed from the analysis.
By comparing the runtime performance of apps debloated by \nameDe with those on the original AOSP, we discerned the differences in resource usage attributed to \nameDe's debloating process.

\textbf{Experiment results.}
\myfig~\ref{fig_rq4} shows the ratio of the resources (i.e., CPU and memory) consumed by the debloated apps and original apps, with the red bars representing the median values.
For DEX code debloating, \nameDe exhibited a median reduction rate of 14.2\% on CPU usage and 0.3\% on memory usage.
This reduction is attributed to \nameDe intercepting the invocation of debloated methods, preventing their code from being compiled and executed.
Regarding the native code debloating, we observed that \nameDe achieved median reduction rates of 28.8\% on CPU usage and 0.7\% on memory usage.
The reduced CPU usage is primarily due to the fact that the debloated library method is zero-filled, thus it will not be executed.
As our zero-filling process does not directly reduce memory usage, the memory usage reduction is primarily because the methods called by the debloated methods were not loaded into memory.

We also observed that in some cases, debloating apps with \nameDe consumed slightly more resources than without debloating. 
One reason is that \nameDe checks if an invoked method belongs to the debloating schema, which introduces additional processing in ART, leading to a slightly higher resource consumption than running the apps on the original Android OS. 
However, in general, \nameDe reduced the consumption of CPU and memory usage compared to running apps without debloating.

\begin{center}
    \vspace{-2ex}
    \begin{mybox}[boxsep=0pt,
	boxrule=1pt,
	left=4pt,
	right=4pt,
	top=4pt,
	bottom=4pt,
	]

    \textbf{Answer for RQ4:}
        The comparison result between running the apps on \nameDe and the original AOSP indicates that debloating apps with \nameDe can reduce the consumption of CPU and memory resources.
    \end{mybox}
    \vspace{-1ex}
\end{center}
\section{Discussion}
\label{sec:discussion}


\subsection{Preventing Potential Ethical Issues}
\nameDe is implemented on an unrooted device, and all the functionalities work without rooting the device.
\nameDe does not inject malicious content into the apps' code except for the simple return instructions (described in \mysec\ref{sec:nativeDebloat}).
While conducting the experiment in \mysec\ref{sec:evaluation}, all the debloating processes were conducted on the local apps without sending any intentional data to the apps' servers or interacting with other app users.
To mitigate ethical concerns during debloating, we filter out security-related methods while generating the debloating schema, e.g., methods of which the names contain keywords like encryption, passwords, etc. 
We also excluded all methods from standard Java and Google APIs.
As a result, \nameDe did not impact the security-related functionalities of the apps.
Moreover, we did not modify the default SELinux policy of the Android system~\cite{gove2016v3spa}, and all implementations followed the default SELinux policy to avoid importing additional vulnerabilities.
Additionally, since the read-only permission for the ContentProvider is not a sensitive permission, it is granted to every app during installation. 
Nevertheless, this permission can be changed to a runtime permission, which requires users' agreement before being granted upon the app launch.

\subsection{Limitations}
\textbf{Threat to validity.}
\nameDe leverages Android's ContentProvider to share the debloating schema, which can prevent adversaries from modifying the schema to prevent any app from being debloated.
However, we assume the input debloating schema is from a trusted source. 
One possible solution is to generate a schema database for commonly used applications and obtain authorization from the developers of Android OS and apps. 
Moreover, apps may attempt to detect \nameDe through querying the ContentProvider of our management app and potentially evade debloating by force-stopping their apps.

Although \nameDe can recover a debloated DEX method by removing it from the debloating schema, debloating a new method could be complex. 
Assume there is a method $m$ that is not initially included in the debloating schema.
After the app has been running for some time, $m$ may have been JIT/AOT compiled, and subsequent invocations of $m$ are directed to the compiled native code, which does not rely on the interpreter.
In that case, \nameDe cannot debloat this method $m$ by simply adding it to the schema and applying the new schema.
To debloat this newly included method, users can either reinstall the app or clear the app's data to eliminate the generated native code before applying the updated debloating schema.

Since \nameDe leverages the same debloating schema as the static debloating tools, both static and dynamic debloating approaches may encounter the same problem, such as missing expected return values of a debloated method, potentially causing app crashes.
However, given that the methods within the schema are rarely used, this situation will not be frequent, and those accidental triggers will be redirected to our graceful termination (see \mysec\ref{sec:DEXDebloat} and \mysec\ref{sec:nativeDebloat}). 
To completely address such issues, enhancing the accuracy of the debloating schema is necessary, which falls outside the scope of \nameDe. In this paper, we assume that \nameDe has a properly generated debloating schema as input.

\textbf{Potential risks of native code debloating.}
In \mysec\ref{sec:nativeDebloat}, during the process of zero-filling the native code in the memory, we temporarily set the memory segment to be both writable and executable, and we immediately remove the writable permission after completing the zero-filling of the native method's code.
Despite our best efforts, there remains a small time window for potential attackers to write and execute in memory. 
To mitigate this risk, one possible solution is to map the library code into a temporary memory space that is writable but non-executable.
After that, we conduct zero-filling on the memory space of each target method, then copy the modified memory content back to the original memory space, which is executable. 
In this paper, our current implementation serves to demonstrate the concept of debloating native methods, and we plan to refine it in the future.

\subsection{Defending code reuse attacks via direct inter-app code invocation}
\label{sec:dici}
\vspace{-0.3cm}
Direct inter-app code invocation (DICI) is a mechanism based on Java reflection and specific API methods provided by the Android framework, allowing one app to invoke methods from another app. 
When \texttt{app A} invokes a method from \texttt{app B}, the code runs within \texttt{app A}'s process. 
Previous research has shown that attackers can exploit DICI to access private data and conceal malicious actions~\cite{gao2020borrowing}, such as retrieving IMEI numbers or sending SMS messages by invoking methods from victim apps.
Lin et al. introduced a cache side-channel attack that infers user behavior by using the DICI mechanism to monitor app-specific methods executed in victim processes~\cite{lin2024peep}.
\nameDe applies to defending against these DICI attacks.
However, unlike other code reuse attacks that target unused code, the victim code may still be used by the victim app.
Therefore, defending against such attacks requires preventing the victim code from being executed by other apps while the victim app can still run it.
Under this scenario, \nameDe can be configured with a special \textit{whitelist} mode, i.e., given a list of methods and the victim app's package name, only the victim app can run the code while all the executions of these methods from all other apps will be blocked.
Traditional static debloating approaches~\cite{jiang_reddroid_2018,liu2023autodebloater,liu2024minimon,pilgun_dont_2020,xdebloat-tang-2022,thung2024towards} cannot achieve this, since they remove unused code so the victim app itself cannot run the code.


\subsection{Possible Alternative Implementation}
\label{sec:alterImplement}
As introduced in \mysec\ref{sec:design}, \nameDe is implemented by customizing the AOSP, which requires end users to flash the customized Android OS on their devices. 
While it is feasible for companies or organizations to uniformly manage their devices with a customized Android OS installed, it presents a challenge for personal users who typically use the OS provided by different phone vendors, such as MIUI from Xiaomi~\cite{MIUI} and One UI from Samsung~\cite{OneUI}. 
This issue can be addressed by collaborating with phone vendors to integrate the functionalities of \nameDe into their customized Android OS.

Since we are not currently cooperating with OS vendors, we explore a possible alternative implementation for dynamic debloating to facilitate use by personal users.
We consider leveraging the Linux Extended Berkeley Packet Filter (eBPF)~\cite{eBPF}, a popular technology that 
can safely and efficiently extend kernel capabilities without requiring changes to the kernel source code.
Previous works have applied this eBPF technique in Android system for various purposes, such as malware detection (e.g., BPFroid~\cite{agman2105bpfroid}) and native code analysis (e.g., NCScope~\cite{zhou2022ncscope}). 
Sifter~\cite{hung2022sifter} mitigates vulnerabilities in security-critical kernel modules in Android by monitoring the system calls using eBPF.
The eBPF is enabled by default on recent Android devices, however, no existing work has utilized eBPF for dynamically debloating Android apps.

The eBPF-based implementation is based on self-customized eBPF programs that enable executing programs in a privileged context within the Linux kernel.
These eBPF programs are loaded during system boot, and can dynamically insert probes into programs and hook the function to execute corresponding operations.
It can hook a kernel instruction by using the kernel probe (kprobe)~\cite{kprobe} or hook user-space programs through user-space probe (uprobe)~\cite{uprobe}.
Additionally, the eBPF program is verified by the kernel versifier before being loaded into the memory, ensuring that the program does not crash the system or access invalid memory addresses, etc.
One can attach eBPF programs to the uprobes or kprobes and collect useful kernel statistics, monitor, and debug.
Different from modifying the source code, dynamically inserting probes requires locating the specific memory address of corresponding programs and the function offsets.
For example, the ART module of the AOSP is compiled to the \textit{libart.so} library, and we need to locate the address of the loaded \textit{libart.so} and the offsets of the functions we would like to monitor.
The specific ideas for an eBPF-based implementation are elaborated in Appendix~\ref{sec:eBPFAppendix}.

Instead, they only need to push the eBPF program and corresponding debloating schema onto the device and restart it to activate the eBPF program. The debloating process will then take effect. 
The main limitation of this eBPF solution is that users may not be able to conveniently change the debloating schema during runtime, as the eBPF program is loaded only during boot.
We line this alternative implementation at the initial stage without conducting extensive experiments, and this approach would be considered as a future extension of this work.

\section{Related Works}
\label{sec:relatedDe}
\textbf{Code reuse attacks in Android apps.}
Although ROP attacks were initially introduced on computer OS~\cite{shacham2004effectiveness}, it has been demonstrated that Android OS is also susceptible to such attacks. 
Existing research on Android apps' code reuse attacks primarily focuses on defending against native code reuse~\cite{raja2017return}, specifically code in native libraries of Android apps and system libraries.
Additionally, Gao et al. analyzed the direct inter-app code invocation among Android apps~\cite{gao2020borrowing}, which can be leveraged to perform malicious attacks by exploiting the code from another app.
Sun et al.~\cite{sun2016blender} proposed Blender, which can self-randomize the address space layout of apps to mitigate the bypassing ASLR protection on Android systems, making it difficult for attackers to identify and collect gadgets for exploitation. 
Unlike previous approaches that rely on memory address randomization, our approach takes a different route to mitigate code reuse attacks in the Android system,  preventing the DEX method from compilation and removing native methods code when loading the native libraries into the memory.

\textbf{Debloating android apps.}
Google has recognized the importance of debloating Android apps and provided solutions from the developers' perspective.
For example, Google provides a static analysis tool, i.e.,  R8, to detect and remove unused DEX code and resources from apps~\cite{shrink--}.
Additionally, Google allows developers to use the App Bundle format so that only the necessary code and resources for a specific device or feature are downloaded~\cite{android--c}.
In academia, researchers have developed a series of approaches to debloat Android apps.
Jiang et al. remove dead code of Android apps based on static analysis \cite{jiang_reddroid_2018}.
Pilgun et al. debloated apps by removing the code that is not executed during the test \cite{pilgun_dont_2020}.
Tang et al. debloated apps at the granularity of Activity, Permission, and Modularity \cite{xdebloat-tang-2022}.
Liu et al. developed a monitor-based framework called MiniMon which debloats apps based on the usage of users~\cite{liu2024minimon}.
Thung et al. constructed partial call graphs to speed up permission-based app debloating~\cite{thung2024towards}.
TaintART~\cite{sun2016taintart} and NDroid~\cite{qian2014tracking} are proposed to dynamically track sensitive information flowing through JNI. 
Unlike previous works, our approach differs in the following aspects: (1) we perform runtime app debloating rather than statically modifying the APK file, (2) we are capable of debloating native library methods in Android apps, not limited to just DEX code.


\textbf{Debloating binary programs.}
Researchers proposed a series of approaches to identify the code to debloat based on static binary analysis.
For example, Agadakos et al. removed the unused code by taking advantage of debug symbols to identify function boundaries, construct library function call graphs, and detect address-taken functions that could be targeted by indirect calls \cite{agadakosNibbler2019}.
Landsborough et al. employed a genetic algorithm in toy programs that disabled features in binaries \cite{landsborough_removing_2015}.
Qian et al. use heuristics to identify unnecessary basic blocks and remove them from the binary \cite{qianRazor2019}.
Ghaffarinia and Hamlen used a similar approach based on training to limit control flow transfers to unauthorized sections of the code \cite{ghaffarinia_binary_2019}.
DamGate~\cite{chen_damgate_2017} rewrites binaries with gates to prevent the execution of unused features, which is most related to our paper. 
As highlighted in prior research~\cite{sun2014nativeguard, qian2014tracking, alam2017droidnative}, the native library code is also susceptible to malicious behaviors and vulnerabilities.
Jucify~\cite{samhi2022jucify} unifies DEX and native code to support comprehensive static analysis of
Android apps, which can be leveraged to identify the native code to be debloated.
JN-SAF~\cite{wei2018jnSAF} is an inter-language static analysis framework to detect sensitive data leaks in Android apps, which combines tho DEX and native code.
Different from their approaches, we do not modify the native library file. Instead, we selectively load native methods during the library loading process.

\section{Conclusion}
\label{sec:conclusion}
In this paper, we propose \nameDe, a late-stage framework for conducting dynamic method-level debloating that empowers users to selectively debloat Android app methods at runtime. 
Different from existing debloating approaches that focus on identifying code to be debloated but debloat the app by modifying the APK, \nameDe offers a dynamic debloating process that preserves the integrity of APKs. 
Under the guidance of a user-defined debloating schema, \nameDe intercepts the invocations and execution of target methods and prevents target code loading into the memory, covering both DEX and native methods.
Our evaluation demonstrates \nameDe's effectiveness in debloating the DEX and native methods in real-world apps, showcasing reduced resource consumption compared to running apps without debloating.   
Furthermore, by debloating DEX and native methods, \nameDe significantly decreases potential Return-Oriented Programming (ROP) gadgets, and mitigates the vulnerabilities in the unused code, diminishing the attack surface.
In addition to modifying the AOSP, we also explore a potential eBPF-based implementation of \nameDe for more accessible and more friendly integration with commercial Android OS developed by third-party vendors.
In the future, \nameDe can be applied to more scenarios, e.g., blocking potential malicious code in Android apps.

\bibliographystyle{plain}
\bibliography{main}

\appendices
\section{Specific eBPF Implementation}
\label{sec:eBPFAppendix}
The specific ideas for an eBPF-based implementation of \nameDe are listed as following:

\textbf{Debloating Schema Configuration.}
By using eBPF, we do not need the ContentProvider to transfer the debloating schema to the Android Runtime. Instead, the eBPF program can read the schema file directly, eliminating the need to grant extra permissions to every installed app for accessing the debloating schema. 
Our management app can still be used to set and modify the debloating schema file. 
The debloating schema is an individual file which is separated from the eBPF program, i.e., users can change the debloating schema file without changing the eBPF program.
However, since the eBPF program is loaded during Android boot, once the user decides to change the debloating schema, she will need to reboot the device after modifying the schema file for the changes to take effect.

\textbf{DEX method debloating via eBPF.}
The DEX method debloating is primarily achieved through the \textit{ART} module~\cite{androidruntime} in the AOSP, which is compiled into the \textit{libart.so} library when building the system image. 
Therefore, we can add uprobes in \textit{libart.so} via eBPF to monitor the invocation of methods. 
If a monitored method belong to the pre-configured debloating schema, we can intercept its invocation in the corresponding handlers in the eBPF program.

\textbf{Native method debloating via eBPF.}
The method debloating of native libraries is primarily achieved by modifying the \textit{bionic} module~\cite{androidbionic} in the AOSP, which compiles into multiple \textit{so} libraries during the system image compilation. 
\nameDe's functionality is implemented in the compiled \textit{libc.so} (for system calls) and \textit{libdl.so} (for loading and linking native libraries).  
By inserting probes into the corresponding functions within these two libraries, we can zero-fill the target code of native methods specified in the schema during the library loading process (as described in Section \ref{sec:nativeDebloat}).

\textbf{Graceful termination.}
The graceful termination in the eBPF-based implementation can follow the approach described in  \mysec\ref{sec:DEXDebloat} and \mysec\ref{sec:nativeDebloat}.
The only difference is that redirection to graceful termination is implemented within the handlers of the eBPF programs.

\section{Related Code and Commands}
\label{sec:codeandCommands}
The following is the command triggering the AOT compilation that compiles the DEX methods into native code (i.e., ARM instructions) based on a profile.
\begin{lstlisting}[
    caption={The command for the profile-based AOT compilation},  
    label={code:AOTcompilation},           
    basicstyle=\footnotesize\ttfamily,           
    numbers=left,                   
    numberstyle=\tiny\color{gray},  
    stepnumber=1,                   
    numbersep=5pt,                 
    frame=single,                   
    framerule=1pt,                  
    rulecolor=\color{black},        
]
adb shell cmd package compile -m speed-profile <package_name>

\end{lstlisting}

The following is an example of defining the read-only permission of a ContentProvider. 
These attributes of the ContentProvider are defined in the Manifest file of an app.
Other apps need to be granted this permission to access this ContentProvider. 

\begin{lstlisting}[
    language=XML,                
    caption={An example of defining read-only permission of a ContentProvider in the Manifest file},  
    label={code:cpPermission},           
    basicstyle=\footnotesize\ttfamily,           
    numbers=left,                   
    numberstyle=\tiny\color{gray},  
    stepnumber=1,                   
    numbersep=5pt,                 
    frame=single,                   
    framerule=1pt,                  
    rulecolor=\color{black},        
]
<provider
 android:name="com.example.mycp"
 android:authorities="com.example.mycp"
 android:exported="true"
 android:readPermission="com.example.mycp.READ"
/>

\end{lstlisting}

\end{document}